\documentclass[twocolumn,superscriptaddress,showpacs,preprintnumbers,amsmath,amssymb,prl]{revtex4}
\usepackage{graphicx}% Include figure files
\usepackage{dcolumn}% Align table columns on decimal point
\usepackage{bm}% bold math

\begin{document}

\title{Solitons in two-dimensional Bose-Einstein condensates}

\author{Shunji Tsuchiya}
\email{tsuchiya@rk.phys.keio.ac.jp} \affiliation{Department of
Physics, Keio University, 3-14-1 Hiyoshi, Kohoku-ku, Yokohama
223-8522, Japan}
\affiliation{CREST(JST), 4-1-8 Honcho, Saitama 332-0012, Japan}
\affiliation{CNR INFM-BEC and Dipartimento di
Fisica, Universit\`a di Trento, I-38100 Povo, Italy}
\author{Franco Dalfovo}%
\affiliation{CNR INFM-BEC and Dipartimento di Fisica, Universit\`a
di Trento, I-38100 Povo, Italy}
\author{Lev Pitaevskii}%
\affiliation{CNR INFM-BEC and Dipartimento di Fisica, Universit\`a
di Trento, I-38100 Povo, Italy}
\affiliation{Kapitza Institute for
Physical Problems, 119334 Moscow, Russia}

\date{March 14, 2008}

\begin{abstract}
The excitations of a two-dimensional (2D) Bose-Einstein condensate
in the presence of a soliton are studied by solving the
Kadomtsev-Petviashvili equation which is valid when the velocity
of the soliton approaches the speed of sound. The excitation
spectrum is found to contain states which are localized near the
soliton and have a dispersion law similar to the one of the stable
branch of transverse oscillations of a 1D gray soliton in a 2D
condensate. By using the stabilization method we show that these
localized excitations behave as resonant states coupled to the
continuum of free excitations of the condensate.
\end{abstract}


\maketitle

Bose-Einstein condensates of ultracold atoms are ideal systems for
exploring matter wave solitons \cite{book}. For most purposes,
these condensates at zero temperature are well described by the
Gross-Pitaevskii (GP) equation \cite{PG} which has the form of a
nonlinear Schr\"odinger equation, where nonlinearity comes from
the interaction between atoms. In one dimension, the GP equation for 
condensates with repulsive interaction admits
solitonic solutions corresponding to a local density depletion,
namely gray and dark solitons. Such solitons have already been
created and observed in elongated condensates with diverse
techniques \cite{Burger}.

Also multidimensional solitons in condensates have attracted much
attention \cite{Brand}. An interesting type of excitation in a two-dimensional (2D)
condensate is represented by a self-propelled vortex-antivortex
pair which is a particular solitonic solution of the GP equation
\cite{Jones,Jones2,Berloff,Berloff2}. In the low momentum limit, 
the relation between the energy $E$ and momentum ${\bf P}$ of the 
soliton \cite{Jones} approaches the dispersion law of Bogoliubov 
phonons, $\epsilon= c q$, from below.  In this limit, when the 
2D soliton moves at a velocity $V$ close to the Bogoliubov sound 
speed $c$, the phase singularities of the vortex-antivortex pair 
disappear and the soliton takes the form of a localized density 
depletion, also called rarefaction pulse. Moreover, if $V$ is close 
to $c$ the GP equation can be rewritten in a simpler form, known as the
Kadomtsev-Petviashvili (KP) equation \cite{KP}.

In a previous short paper \cite{Tsuchiya}, we already presented
some preliminary results on the dynamics and stability of this 2D
soliton. In particular, we studied the excitations of the
condensate in the presence of the soliton by linearizing the KP
equation around the stationary solution. By looking at the shape
of the eigenfunctions we found excitations localized near the
soliton, having shape and dispersion law similar to those of the
transverse oscillations of a 1D gray soliton in a 2D condensate.
In this work we present a more systematic analysis. We use a
stabilization method in order to obtain a better determination of
the dispersion law of the localized states. Moreover, the same
method allows us to visualize the coupling between the localized
states and the free states, i.e., the Bogoliubov phonons of the
uniform condensate.

Let us first summarize the derivation of the KP equation. We
consider a 2D condensate with a soliton moving at a constant
velocity $V$ in the $x$-direction. If the density at large
distances is $n_{\infty}$, one can define the healing length
$\xi=\hbar/[2mgn_\infty]^{1/2}$, where $g$ is the mean-field
coupling constant and $m$ is the mass of the bosons.  One can also
introduce the dimensionless variables $x \to \xi x$, $y \to \xi y$,
and $t \to mt \xi^2/\hbar$, the normalized order parameter $\Psi
\to \sqrt{n_\infty} \Psi$, and the velocity $U=m\xi V/\hbar=
V/(c\sqrt{2})$.  In the frame moving with the soliton, the GP
equation is \cite{Jones}
%%%%%%%%%%%%%
\begin{eqnarray}
2i \frac{\partial \Psi}{\partial t} =
-\nabla^2 \Psi + 2iU \frac{\partial}{\partial x}\Psi +
\left(|\Psi|^2-1\right)\Psi \; .
\label{GP}
\end{eqnarray}
%%%%%%%%%%%%%
The order parameter can be written in the form
$\Psi=n^{1/2}\exp[iS]$. The equations for the phase and the
density thus become
%%%%%%%%%%%%%
\begin{eqnarray}
\frac{\partial S}{\partial t}&=&\frac{1}{2}\frac{\nabla^2 n^{1/2}}{n^{1/2}}
-\frac{1}{2}\left(\nabla S\right)^2-\frac{1}{2}(n-1)
+U\frac{\partial S}{\partial x}\;\; ,\label{eqmotion}\\
\frac{\partial n}{\partial t}&=&-(\nabla n)\cdot(\nabla S)-n(\nabla^2 S)
+U\frac{\partial n}{\partial x}\;\; .\label{contieq}
\end{eqnarray}
%%%%%%%%%%%%%
When $V$ is close to $c$, one has $U \simeq 1/\sqrt{2}$. Let us
introduce a small parameter $\varepsilon\equiv\sqrt{1-2U^2}$, so
that $U \simeq 1/\sqrt{2} - 1/(2\sqrt{2}) \varepsilon^2$, and
expand the density and the phase in this form
 %%%%%%%%%%%%%
 \begin{eqnarray}
 n&=&1 - \varepsilon^2f+\dots\; \;,  \label{density}\\
 S&=&\varepsilon s +\dots\; \; .
 \end{eqnarray}
%%%%%%%%%%%%%
To the lowest order in $\varepsilon$, the GP equation gives
$\partial s / \partial x = - \varepsilon f/\sqrt{2}$ and
%%%%%%%%%%%%%
\begin{eqnarray}
\frac{\partial}{\partial \tilde{x}}\left(\frac{\partial f}{\partial \tilde{t}}
+6f\frac{\partial f}{\partial \tilde{x}}+\frac{\partial^3 f}{\partial \tilde{x}^3}
\right) = \frac{\partial^2 f}{\partial  \tilde{y}^2} \; \; ,
\label{KP}
\end{eqnarray}
%%%%%%%%%%%%%%
where we have introduced the stretched variables $\tilde x= -\varepsilon x +
\varepsilon^3t/(2\sqrt{2})$, $\tilde y= \varepsilon^2 y/\sqrt{2}$
and $\tilde t=\varepsilon^3 t/(4\sqrt{2})$. Equation (\ref{KP}) is known as
the Kadomtsev-Petviashvili (KP) equation \cite{KP}. Hereafter we will omit the
tilde in the stretched variables.

The KP equation (\ref{KP}) admits a stationary solution of the
form
%%%%%%%%%%%%%%
\begin{eqnarray}
f_0(x-2t)= {\rm sech}^2 \left[ (x-2t)/\sqrt{2} \right] \;\; .
\label{KdV}
\end{eqnarray}
%%%%%%%%%%%%%%
which is independent of $y$ and is a limiting form of a 1D gray 
soliton \cite{Tsuzuki}. The linear stability of this gray soliton in 
two dimensions was studied in \cite{Zakharov,Alexander}. One can look for 
transverse fluctuations propagating along $y$ of the form $f(x,y,t) =
f_0(x-2t)+\psi(x-2t)e^{i(ky-\omega t)}$ and solve the KP equation
up to terms linear in $\psi$. One obtains the dispersion law
%%%%%%%%%%%%%%
\begin{eqnarray}
\omega^2=\frac{16}{3\sqrt{3}}k^2 \left( k-\frac{\sqrt{3}}{2}
\right) \;\; . \label{gsolidisp}
\end{eqnarray}
%%%%%%%%%%%%%%
The frequency $\omega$ is real for $k \ge \sqrt{3}/2$ and imaginary 
for $k < \sqrt{3}/2$. The gray soliton (\ref{KdV}) is thus unstable:
long wavelength transverse oscillations can grow exponentially
(snake instability). We note also that for $k\gg 1$ the stable
branch of excitations behaves as $\omega \propto k^{3/2}$,
similarly to the dispersion of capillary waves on the surface of a
liquid.

The KP equation  (\ref{KP}) admits also a 2D solitonic solution of
the form \cite{Manakov}
%%%%%%%%%%%%%%
\begin{eqnarray}
f_0(x-2t,y) =\frac{ 4 [ \frac{3}{2} + 2y^2- (x-2t)^2]}{
[\frac{3}{2} + 2y^2 + (x-2t)^2 ]^2} \; \; ,
\label{2Dsoliton}
\end{eqnarray}
%%%%%%%%%%%%%%
which corresponds to the $V \to c$ limit of the rarefaction pulse
\cite{Jones}. The function $f_0$ is plotted in Fig.~1. Differently
from the 1D gray soliton (\ref{KdV}), the solution
(\ref{2Dsoliton}) decays algebraically in all directions.

%%%%%%%%%%%%%%
\begin{figure}
\centerline{\includegraphics[width=7cm]{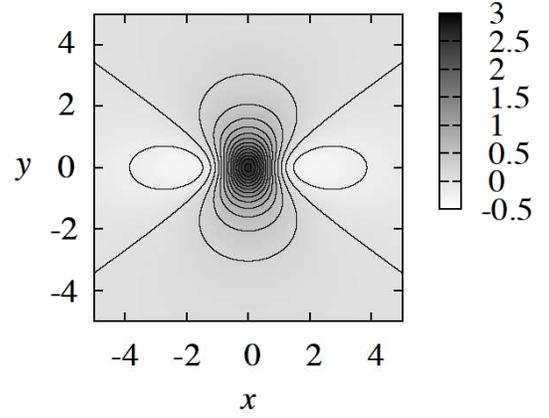}}
\caption{Solitonic solution (\protect\ref{2Dsoliton}) of the KP
equation  (\ref{KP}) as a function of the stretched dimensionless lengths $x$
and $y$ (see text for definitions). For a given value of the small
parameter $\varepsilon$, the density distribution of the
condensate is given by $n=1 - \varepsilon^2 f$ and the ratio
between the soliton widths along $x$ and $y$, in dimensional
units, scales as $\varepsilon$.  } \label{fig1}
\end{figure}
%%%%%%%%%%%%%%

In the units of the GP equation (\ref{GP}), the density profile of
the soliton is found by inserting Eq.~(\ref{2Dsoliton}) into
Eq.~(\ref{density}). Its size along $x$ and $y$ is of the order of
$1/\varepsilon$ and $1/\varepsilon^{2}$, respectively.
To the first order in $\varepsilon$, the momentum $P$ and the
energy $E$ per unit length are \cite{Jones}
%%%%%%%%%%%%%%
\begin{eqnarray}
P= \frac{8\pi\sqrt{2}}{3} \hbar n_{\infty} \xi\varepsilon \; \; ,
\; \; E= \frac{8\pi}{3m} \hbar^2n_{\infty}\varepsilon = cP \; .
\label{EP}
\end{eqnarray}
%%%%%%%%%%%%%%
As it must be, the soliton has a sound-like dispersion in the limit
where the KP equation is valid, that is, when $P \to 0$.

Now we consider density fluctuations of the form
$f(x,y,t)=f_0(x-2t,y)+\psi(x-2t,y)e^{-i\omega t}$. Inserting this
expression into Eq.~(\ref{KP}) and keeping terms linear in $\psi$,
we find
%%%%%%%%%%%%%%
\begin{eqnarray}
\frac{\partial^4 \psi}{\partial x^4}+6\frac{\partial^2}{\partial
x^2} \left( f_0\psi\right)-2\frac{\partial^2 \psi}{\partial x^2}
-\frac{\partial^2 \psi}{\partial y^2}=i \omega \frac{\partial
\psi}{\partial x} \; \; . \label{linKP2D}
\end{eqnarray}
%%%%%%%%%%%%%%
It is convenient to introduce the function $\phi(x,y)$ such that
$\psi= \partial ^2 \phi /\partial x^2$. By integrating
Eq.~(\ref{linKP2D}) twice in $x$, one has 
%%%%%%%%%%%%%%
\begin{eqnarray}
\frac{\partial^4 \phi}{\partial x^4}
+ 6f_0\frac{\partial^2 \phi}{\partial x^2}
-2\frac{\partial^2 \phi}{\partial x^2}
-\frac{\partial^2 \phi}{\partial y^2}
=i\omega \frac{\partial \phi}{\partial x} \; \;.
\label{linKP2D2}
\end{eqnarray}
%%%%%%%%%%%%%%
We numerically solve this equation in a box of size $L_x\times
L_y$. The function $\phi$ is expanded in plane waves with periodic
boundary conditions: $\phi(x,y) = \sum_{\nu,\mu} \phi_{\nu \mu}
\chi_{1,\nu}(x)\chi_{2,\mu}(y)$, with $\chi_{1,\nu}(x) =
L_x^{-1/2} e^{i2\pi \nu x/L_x}$ for $|\nu| \le l_x \;(\nu\neq 0)$.
Concerning $\chi_{2,\mu}(y)$ we note that Eqs.~(\ref{linKP2D}) and
(\ref{linKP2D2}) are invariant for $y\to -y$. Therefore the
function $\phi$ is either an even or odd function of $y$. If it is
even, then one can take $\chi_{2,\mu}(y)= (2/L_y)^{1/2} \cos (2\pi
\mu y/L_y)$ for $1 \le \mu \le l_y$ and
$\chi_{2,0}(y)=(1/L_y)^{1/2}$. If it is odd, one can take
$\chi_{2,\mu}(y)= (2/L_y)^{1/2}\sin (2\pi \mu y/L)$ for $1 \le \mu
\le l$. One thus obtains the following matrix equation
%%%%%%%%%%%%%%
\begin{eqnarray}
\left(-q_x^3-2q_x-\frac{q_y^2}{q_x} \right)\phi_{\nu \mu}+6\sum_{\nu^\prime, \mu^\prime}
\frac{(q_x^\prime)^2}{q_x} M_{\nu \mu , \nu^\prime \mu^\prime} \phi_{\nu^\prime \mu^\prime} \nonumber\\
=\omega \phi_{\nu \mu},\label{mequation}\\
M_{\nu \mu , \nu^\prime \mu^\prime} = \! \int_{-L_x/2}^{L_x/2}\!
dx \! \int_{-L_y/2}^{L_y/2} \! dy \
\chi_{1,\nu}^\ast\chi_{2,\mu}^\ast f_0 \chi_{1,\nu^\prime}\chi_{2,
\mu^\prime}, \label{mm}
\end{eqnarray}
%%%%%%%%%%%%%%
where $q_x=2\pi \nu/L_x$, $q_y= 2\pi \mu/L$.
The size of the matrix is $N\times N$, where $N=2l_x(l_y+1)$ for
$\phi$ even, and $N=2l_xl_y$ for $\phi$ odd.

\begin{figure}
\centerline{\includegraphics[width=8cm]{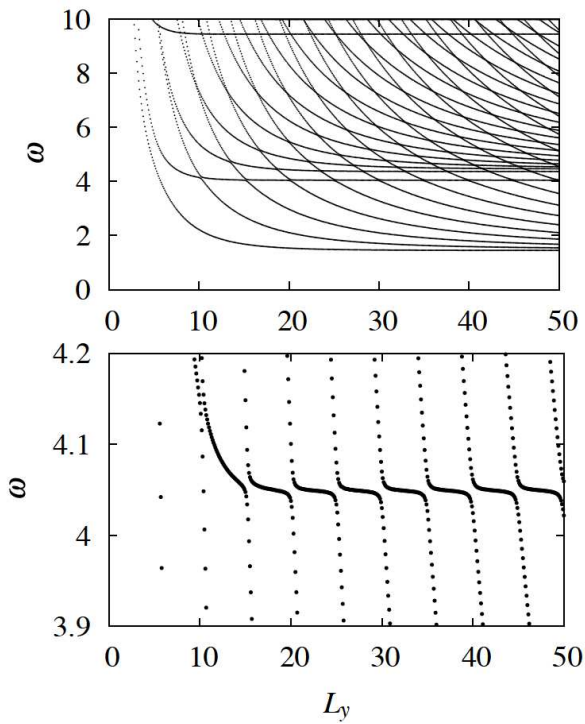}}
\caption{Eigenfrequencies $\omega$ of Eq.~(\ref{mequation}) as a
function of $L_y$ for $L_x=10$.  The lower plot is a magnification
of the upper one around $\omega=4$. All quantities are
dimensionless. }
\label{fig2}
\end{figure}

\begin{figure}
\centerline{\includegraphics[width=8cm]{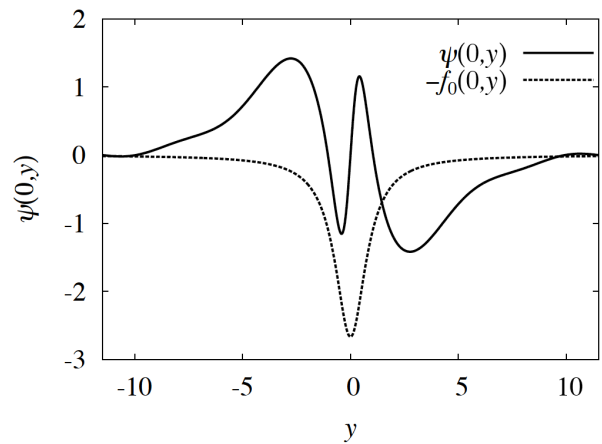}} \caption{Wave
function $\psi(0,y)$ (solid line) of an excited state localized
near the soliton. The soliton profile,  $-f_0(0,y)$, is shown as a
dashed line. The excited state has frequency $\omega=4.05$ and the
box size is $L_x=10$ and $L_y=23$. All quantities are dimensionless.} \label{fig3}
\end{figure}

By solving Eq.~(\ref{mequation}), we find that all eigenvalues are
real and positive.  This is consistent with the stability of the
2D soliton \cite{Berloff, Kuznetsov}. As already discussed in
\cite{Tsuchiya}, among the eigenvectors we find states which are
localized near the soliton. In order to better identify these
states and explore their coupling with the free, unbound states we
use a stabilization method \cite{Mandelshtam}. In this method, the
eigenvalues are calculated repeatedly for different box sizes
$L_y$ and $L_x$. Once the frequency is plotted as a function of
the box size, the localized states are identified as those having
a dispersion which becomes flat for large boxes, larger than the
typical size of the corresponding bound state. These states are
immersed into a bath of unbound states. The latter are
characterized by the number of nodes of the eigenfunctions in the
$x$ and $y$ directions and, due to the finite box, they appear as
series of discrete branches in the stabilization diagram. The
dispersion laws of the unbound states, as well as their dependence
on $L_y$ and $L_x$, can be estimated analytically by inserting
$f_0 =0$ into the equations (\ref{mequation}) and (\ref{mm}). A 
coupling between bound (resonant) states and unbound (free) states 
can be seen in the stabilization diagram in the form of avoided 
crossings due to non-zero matrix elements connecting the two types 
of states \cite{Mandelshtam}. A typical diagram is shown in 
Fig.~\ref{fig2}, where we plot the eigenfrequencies of 
Eq.~(\ref{mequation}) as a function of $L_y$ for $L_x=10$. Flat 
dispersions and avoided crossing are clearly visible. An example 
of localized state is shown in Fig.~\ref{fig3}, where we plot 
the $x=0$ section of the wave function of one of the bound states 
of Fig.~\ref{fig2} together with the soliton profile.

%%%%%%%%%%%%%%
%%%%%%%%%%%%%%

%%%%%%%%%%%%%%
%%%%%%%%%%%%%%

\begin{figure}
\centerline{\includegraphics[width=8cm]{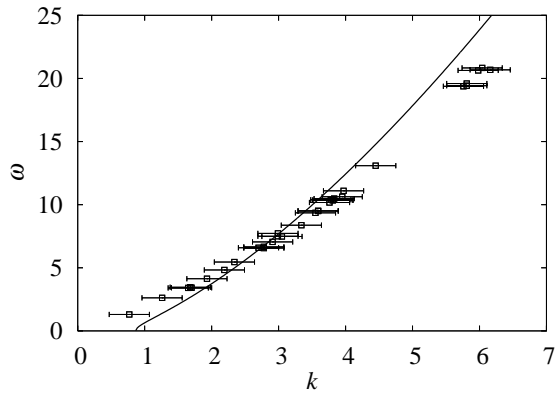}}
\caption{Frequency of lowest excited states of the 2D soliton.
Each point corresponds to an eigenvalue of the matrix equation
(\ref{mequation}), whose eigenfunction $\psi$ is localized in the
soliton region and exhibits oscillations along $y$ with wave vector
$k$.  The solid line is the dispersion law of the stable branch of
excitations of a 1D gray soliton in two dimensions, Eq.~(\ref{gsolidisp}). All
 quantities are dimensionless.}
\label{fig4}
\end{figure}

In the soliton region the wave function of the excited states
exhibits oscillations in the $y$ direction whose wave vector $k$
can be easily estimated. It is interesting to plot the
eigenfrequency $\omega$ of the bound states, extracted from the
stabilization diagram, as a function of $k$, extracted from the
shape of the corresponding wave functions. This is done in
Fig.~\ref{fig4} for the bound states having the lowest number of
nodes in the $x$ directions. Our results are shown as points with
error bars, where the error bars are of the order of the inverse
of the size of the soliton along $y$. In the same figure, the
dispersion law (\ref{gsolidisp}) of the stable branch of
excitations of a 1D gray soliton is plotted. The spectrum is very
similar. This reflects the fact that a 2D soliton moving with
velocity close to $c$ is very elongated in the $y$ direction (the 
ratio between the widths along $x$ and $y$, in dimensional units, 
is proportional to the small parameter
$\varepsilon$) and its density distribution is indeed similar to
that of a 1D gray soliton. Differently from a 1D gray soliton the
2D soliton has a discrete spectrum of bound states due to its
finite length. This finite length also implies an ``infrared cutoff"
in the spectrum of the bound states, so that the long wavelength
transverse oscillations of the infinite 1D soliton, which cause
its snake instability, are not present in the spectrum of the 2D
soliton.

The occurrence of a coupling between bound and unbound states,
which is visible in the avoided crossings in the stabilization
diagram, is worth stressing. The width of the avoided crossings
is directly related to the lifetime of the bound (resonant) states 
associated with their decay into Bogoliubov sound modes. For
typical resonant states, like the one in Fig.~\ref{fig2}, we 
find a width of the resonance, $\Delta \omega$, of the order of 
$5 \times 10^{-3}$ \cite{note}. In \cite{Tsuchiya} we noticed that
the possible existence of bound states with infinite lifetime and
dispersion law $\sim k^{3/2}$ could affect the thermodynamics of 
a 2D condensate. The present analysis suggests that the bound 
states of the 2D soliton have a finite lifetime, hence making the 
problem more complex. 

%%%%%%%%%%%%%%
%%%%%%%%%%%%%%

In conclusion, a rarefaction pulse is an interesting solitonic
excitation of a 2D condensate, which moves at velocity close 
to the Bogoliubov sound and is the low energy counterpart of a
self-propelled vortex-antivortex pair. Its momentum and energy 
were calculated in Ref. \cite{Jones}. The stability of the 
2D rarefaction pulse in the KP limit was derived in \cite{Kuznetsov}
from the boundness of the Hamiltonian.  It was also  investigated 
in \cite{Jones2}, by means of both analytic arguments 
and numerical simulations. More recent numerical results on the 
linear stability have been reported in \cite{Berloff}.
However, a detailed calculation of the spectrum of the 2D condensate 
in the presence of the soliton has not yet been performed. In this 
work we have presented the results of such a calculation, including 
an analysis of resonant states based on a stabilization method. 
We have found localized states which closely resemble the stable 
branch of excitations of a 1D gray soliton. In the stabilization 
diagram, these states appear as resonant states coupled to the bath 
of unbound Bogoliubov phonons. We think that these results can be of 
interest for the investigation of ultracold bosonic gases in disk-shaped 
confining potentials, where vortex pairs are known to play a crucial 
role \cite{Hadzibabic}. The observation of solitons in these systems would 
represent a nice manifestation of nonlinear dynamics in low dimensional
superfluids.

\bigskip

We thank C. Tozzo, C. Lobo, P. Pedri, N. Prokof'ev, B. Svistunov, and N. Hatano for
useful discussions.


\begin{thebibliography}{}
\bibitem{note} A more precise determination of the width and lifetime of the 
resonant states in our 2D geometry would require larger boxes and hence larger 
matrices, beyond the capability of our computation and the scope of this work.
\end{thebibliography}
\end{document}